\documentclass[twocolumn]{aastex631}

\usepackage[T1]{fontenc}
\usepackage{amssymb}
\usepackage{amsmath}
\usepackage{physics}
\usepackage{physunits}
\usepackage{bm}
\let\tablenum\relax

\newcommand{\fqty}[2]{\qty(\frac{#1}{#2})}

\newcommand{\hill}{\textrm{H}}
\newcommand{\kep}{\textrm{K}}

\begin{document}

\title{Oligarchic growth of protoplanets in planetesimal rings}

\author[0009-0007-0041-9222]{Yuki Kambara}
\affiliation{Department of Astronomy, The University of Tokyo, 7-3-1 Hongo, Bunkyo-ku, Tokyo 113-0033, Japan
}

\author[0000-0002-5486-7828]{Eiichiro Kokubo}
\affiliation{Center for Computational Astrophysics, National Astronomical Observatory of Japan, 2-21-1 Osawa, Mitaka, Tokyo 181-8588, Japan}

\begin{abstract}
In the standard planet formation scenario, planetesimals are assumed to form throughout the protoplanetary disk and to be smoothly distributed in the radial direction except for the snowline.
Planetesimal growth has been investigated using this assumption, and the oligarchic growth model is widely accepted.
However, recent simulations of gas and dust evolution have shown that planetesimals form only in radially limited locations---such as at gas pressure bumps and snowlines---and are concentrated in ring-like regions.
When planetesimals are distributed in a ring-like region, scattered ones leak from the ring edge, resulting in planetesimal diffusion.
To investigate protoplanet growth in expanding planetesimal rings, we perform a series of {\em N}-body simulations.
In all the simulations, protoplanet growth is well explained by oligarchic growth, while the ring width expands due to planetesimal scattering by the protoplanets.
Massive protoplanets tend to form near the ring center, and protoplanets that form far from the ring center are less massive than those in the center.
The scaled orbital separations depend on neither the initial ring width nor on the total mass, and they are consistent with estimates based on the oligarchic growth model and the diffused planetesimal distribution.
The width of the expanded planetesimal ring does not depend on its initial width, but it does depend on its total mass.
The maximum mass of protoplanets depends strongly on the total ring mass and weakly on its initial width.
\end{abstract}

\section{Introduction} \label{sec:intro} 

In the standard planet formation scenario \citep[e.g.,][]{standardscenario_Hayashi+1985prpl.conf.1100H,Kokubo+2012PTEP.2012aA308K}, the formation of terrestrial planets is divided into three main stages.
The first stage is the formation of km-sized planetesimals from micron-sized dust in a protoplanetary disk \citep[e.g.,][]{Youdin_Goodman_2005ApJ...620..459Y, Drazkowska+2014A&A...572A..78D, Morbidelli+22, Izidoro+2022NatAs...6..357I, Hyodo+2022A&A...660A.117H}.
Once planetesimals have formed, they orbit around the central star, and their orbits change due to mutual gravitational interactions and gas drag.
The second stage involves the collision and merger of planetesimals, which results in the formation of protoplanets \citep[e.g.,][]{Wetherill+1989Icar...77..330W, KI00, KI02,Jason+2023Icar..39615497W}.
The third stage involves the giant impacts among the protoplanets and the formation of terrestrial planets 
\citep[e.g.,][]{Kokubo+2006ApJ...642.1131K,Kokubo&Genda2010ApJ...714L..21K,Chambers2001Icar..152..205C,Hansen09,Izidoro+2022NatAs...6..357I}.
Some part of a planet's mass can be also gained by the accretion of cm-sized pebbles falling toward the central star \citep[e.g.,][]{Ormel&Klahr_2010A&A...520A..43O, Lambrechts+2019A&A...627A..83L}.

In the standard scenario, planetesimals are assumed to be formed throughout the protoplanetary disk and to be distributed smoothly except for the snowline \citep[e.g.,][]{Hayashi_MMSN_1981PThPS..70...35H,standardscenario_Hayashi+1985prpl.conf.1100H}.
This assumption is currently being revised, however, as research on protoplanetary disks and planetesimal formation progresses.
The latest disk models have shown that planetesimals can form at radially limited locations and can be distributed in ring-like regions \citep{Morbidelli+22,Izidoro+2022NatAs...6..357I,Hyodo+2022A&A...660A.117H}.
Recent ALMA observations have found numerous protoplanetary disks with ring-gap structures \citep[e.g.,][]{ALMA+2015ApJ...808L...3A,PPVII_RingStructure2023ASPC..534..423B}.
Moreover, several studies have shown that an initial concentration of solid masses in ring-like regions is favorable for reproducing the solar system \citep[][]{Hansen09, Izidoro+2022NatAs...6..357I, Morbidelli+22}.
In addition, simulations of the giant impact stage have shown that protoplanets distributed in ring-like regions can reproduce the mass distribution of the terrestrial planets in the solar system \citep{Hansen09, Izidoro+2022NatAs...6..357I}.
It has also been pointed out that the isotopic dichotomy of the solar system can be explained by planetesimal rings that formed in radially distinct locations \citep{Izidoro+2022NatAs...6..357I, Morbidelli+22}.

The growth of planetesimals has been investigated in detail, assuming that they follow a smooth distribution.
In planetesimal disks, the larger planetesimals initially grow faster \citep["Runaway growth;" e.g.,][]{Wetherill+1989Icar...77..330W, KI96+1996Icar..123..180K}.
After protoplanets grow large enough to excite the planetesimals' random velocities, they start to grow in an orderly fashion \citep{IdaMakino1993Icar..106..210I}.
Several dominant protoplanets with similar masses grow while maintaining their orbital separations due to orbital repulsion \citep{KI98, KI00, KI02}.
This process is called oligarchic growth.

When planetesimals are distributed smoothly throughout a disk, the timescale for planetesimal diffusion due to mutual gravity is far longer than the growth timescale \citep[][]{OhtsukiTanaka2003Icar..162...47O}.
Therefore, the global surface density distribution of solid bodies remains almost constant during their evolution \citep[][]{KI02}.
Conversely, when planetesimals are distributed in a narrow ring, planetesimal-planetesimal scattering is expected to cause planetesimals to leak out from the ring edges.
As a result, the ring width expands, and the planetesimal surface density decreases.
This process of planetesimal diffusion may be critical to the planetesimal accretion process because the growth rate and isolation mass of protoplanets in the oligarchic growth stage depend on the planetesimal surface density \citep[e.g.,][]{KI00,KI02}.
Consequently, oligarchic growth in an expanding planetesimal ring is expected to differ from the standard model.
Although the growth of planetesimals in a ring has been studied in some previous works {\citep[][]{Hansen09,Walsh+11Mars_2011Natur.475..206W,Deienno+2019ApJ...876..103D,Jason+2023Icar..39615497W,Batygin&Morbidelli_2023NatAs...7..330B}}, they used superparticles for the planetesimals or considered only the the solar system formation.
The growth of realistic-sized planetesimals ($\lesssim 10^{23}\gm $) and its dependence on ring parameters has yet to be investigated in detail.

To construct an extended oligarchic growth model that takes into account planetesimal diffusion in planetesimal rings, in this study, we investigate the protoplanet growth in a planetesimal ring and its dependence on the initial conditions.
Specifically, we perform a series of {\em N}-body simulations of planetesimal accretion in a ring.
To explore the dependence of protoplanet properties on the initial conditions,
we varied the initial ring width and the total mass systematically.

In Section \ref{sec:model}, we briefly summarize the standard oligarchic growth model \citep{KI98,KI00,KI02}. 
Our simulation setup is explained in Section \ref{sec:method}. 
Section \ref{sec:results} presents our simulation results and discusses the dependence on initial conditions.
Section \ref{sec:summary} is devoted to a summary and discussions.
\section{Oligarchic growth of protoplanets in planetesimal disks}\label{sec:model}

Once the mass ratio of planetesimals becomes sufficiently large, {the excitation of their random velocity (i.e., their eccentricity $e$ and inclination $i$)} is dominated by viscous stirring by the largest planetesimals (protoplanets).
In this stage, larger protoplanets have a lower growth rate than smaller ones because the higher relative velocity between a protoplanet and the excited planetesimals diminishes the gravitational focusing.
In this section, we summarize the oligarchic growth model of protoplanets \citep{KI98,KI00,KI02}.
{Then, we develop it further to compare the model with the results of our {\em N}-body simulations.}
Hereafter, we consider equal-mass planetesimals with mass $m$ and protoplanets with mass $M$.

In the oligarchic growth stage, the planetesimals' random velocity increases due mainly to viscous stirring by the protoplanets.
The planetesimals' typical random velocity is $v \simeq \langle e^2\rangle^{1/2} v_\kep$, where $v_\kep$ is the Kepler circular velocity, and $\langle e^2\rangle^{1/2}$ is the {root-mean-square (RMS)} eccentricity.
At the semimajor axis $a$, the timescale for exciting the planetesimals' random velocities is given by 
\citep{IdaMakino1993Icar..106..210I}
\begin{eqnarray}
    T_{\rm VS} &\simeq& 
    \frac{v^3}{\pi G^2 n_MM^2\ln \Lambda},
    \label{timescale::viscous_protoplanets}
\end{eqnarray}
where $G$ is the gravitational constant, and $\ln \Lambda$ is the Coulomb logarithm.
The protoplanets' number density is
\begin{equation}
    n_M \simeq \frac{1}{2\pi ab2ai},
\end{equation}
where $b$ is the orbital separation between adjacent protoplanets.

When a planetesimal moves in the gas disk, its random velocity is damped by gas drag.
In a disk with gas density $\rho_{\rm gas}$, the timescale for random velocity damping due to the gas drag is given by \citep{Adachi+1976PThPh..56.1756A}
\begin{eqnarray}
    T_{\rm GD} &\simeq& 
    \frac{2m}{C_{\rm D} \pi r^2 \rho_{\rm gas} u},
\label{timescale::gasdrag}
\end{eqnarray}
where $C_{\rm D}$ is the drag coefficient, $r$ is the planetesimal's radius, and $u$ is the relative velocity between the planetesimal and the gas.

Hereafter, we use the scaled eccentricity $\langle \Tilde{e}^2 \rangle^{1/2} \equiv \langle e^2 \rangle^{1/2}/h$, where $h\equiv [(M+m)/3M_\star]^{1/3} \simeq (M/3M_\star)^{1/3} $ is the reduced mutual Hill radius between a protoplanet and a planetesimal, where $M_\star$ is the stellar mass.
In addition, we scale the orbital separation between the protoplanets as $\Tilde{b}\equiv b/r_{\rm H}=(a_{\rm out}-a_{\rm in})/r_{\rm H}$, where $a_{\rm out}$ and $a_{\rm in}$ are the semimajor axes of the inner and outer protoplanets and $r_{\rm H}\equiv a (2M/3M_\star)^{1/3} $ is their mutual Hill radius.
By equating eqs. \eqref{timescale::viscous_protoplanets} and \eqref{timescale::gasdrag}, we obtain the equilibrium eccentricity
\begin{eqnarray}
{\langle\Tilde{e^2}\rangle}^{1/2} &=& \qty(\frac{6 \ln\Lambda}{C_{\rm D}\pi})^{1/5} \qty(3\pi^2)^{1/15} 
\nonumber\\
&&\times
m^{1/15} \rho_{\rm p}^{2/15} 
\Tilde{b}^{-1/5}
\rho_{\rm gas}^{-1/5} a^{-1/5}\nonumber
\\
&\simeq& 5.3
\qty(\frac{m}{10^{23}\gm})^{1/15}
\qty(\frac{\rho_{\rm p}}{2\gm\cm^{-3}})^{2/15} \nonumber\\
&&\times
\qty(\frac{\Tilde{b}}{10})^{-1/5}
\qty(\frac{C_{\rm D}}{1})^{-1/5}
\nonumber\\
&&\times
\qty(\frac{\rho_{\rm gas}}{2\times10^{-9}\gm\cm^{-3}})^{-1/5}
\qty(\frac{a}{1\au})^{-1/5},
\label{equilibllium_eccentricity}
\end{eqnarray}
where $\rho_{\rm p}$ is the planetesimals' mass density, and we used $u\sim v \sim \langle e^2\rangle^{1/2} v_{\rm K}$, $\langle e^2 \rangle^{1/2} =2 \langle i^2 \rangle^{1/2}$ and $\ln\Lambda=3$ in the same way as \cite{KI02}.

The growth rate of a protoplanet with mass $M$ and radius $R$ in a swarm of planetesimals with surface density $\Sigma$ is given by
\citep[e.g.,][]{IdaMakino1993Icar..106..210I}
\begin{equation}
    \dv{M}{t} \simeq \pi \Sigma \frac{2GMR}{\langle e^2 \rangle a^2 \Omega},
\end{equation}
where $\Omega$ is the Kepler angular velocity of the protoplanet.
In this case, the growth timescale is 
\begin{equation}
    T_{\rm grow} \equiv \frac{M}{\dv*{M}{t}} \simeq  \frac{\langle e^2 \rangle a^2 \Omega}{2\pi GR\Sigma}. \label{timescale::growth}
\end{equation}

In a planetesimal disk, protoplanets maintain their orbital separations by orbital repulsion \citep{KI+1995Icar..114..247K}.
The typical orbital separation is determined by the balance between expansion due to protoplanet-protoplanet scattering and the increase in the Hill radius due to protoplanet growth.
The timescale of orbital repulsion is \citep{PetitHenon_1986Icar...66..536P,Hasegawa_and_Nakazawa_repulsion_1990A&A...227..619H,KI98}
\begin{eqnarray}
    T_{\rm repel} \equiv \frac{\Tilde{b}}{{\dd \Tilde{b}}/{\dd t}} \simeq \frac{\Tilde{b}^5}{7h\Omega}. \label{timescale::repelsion}
\end{eqnarray}
Conversely, a protoplanet's Hill radius grows with the typical timescale $T_{r_{\rm H}}=3T_{\rm grow}$.
By equating $T_{\rm repel}$ and $T_{r_{\rm _H}}$, \cite{KI98} estimated the typical orbital separation to be
\begin{eqnarray}
    \Tilde{b}_{\rm typical} &=&
    \fqty{7}{\pi}^{1/5} \fqty{\pi}{3}^{1/15} 
     M^{2/15}
     \langle \Tilde{e}^2\rangle ^{1/5}  
    \Sigma^{-1/5} 
    a^{-1/5} \rho_{\rm p}^{1/15}
    \nonumber\\ 
    &\simeq& 10
    \qty(\frac{\langle \Tilde{e^2} \rangle^{1/2} }{5})^{2/5}
    \qty(\frac{M}{10^{26}\gm})^{2/15}
    \qty(\frac{a}{1\au})^{-1/5}
    \nonumber\\
&&\times
\qty(\frac{\Sigma}{10\gm\cm^{-2}})^{-1/5}
    \fqty{\rho_{\rm p}}{2\gm\cm^{-3}}^{1/15},
    \label{eq::separation}    
\end{eqnarray}
and they confirmed that this estimate matches the results of {\em N}-body simulations well.

The isolation mass of a protoplanet in a planetesimal disk is \citep{KI02}
\begin{eqnarray}    
    M_{\rm iso} &\simeq& 2\pi ab\Sigma \nonumber\\
    &=& (2\pi)^{3/2}\qty(\frac{2}{3})^{1/2}M_\star^{-1/2} a^3\Tilde{b}^{3/2} \Sigma^{3/2}\nonumber\\
    &\simeq& 0.16 \qty(\frac{\Sigma}{10\gm\cm^{-2}})^{3/2}
    \nonumber\\
&&\times
\qty(\frac{b}{10r_\hill})^{3/2}
    \qty(\frac{a}{1\au})^{3} M_\oplus,
    \label{eq:isolationmass}
\end{eqnarray}
when the planetesimal surface density $\Sigma$ is constant.

The equilibrium eccentricity depends on the typical orbital separation $\Tilde{b}$, and vice versa.
Using eqs. \eqref{equilibllium_eccentricity} and \eqref{eq::separation}, we can eliminate this dependence to obtain
\begin{eqnarray}
    \Tilde{b} &\simeq& C_b^{25/27}C_e^{10/27} 
    \rho_{\rm p}^{1/9}\rho_{\rm gas}^{-2/27} 
    m^{2/81} M^{10/81}
    a^{-7/27} \Sigma^{-5/27}\nonumber\\
    &\simeq& 10
    \qty(\frac{m}{10^{23}\gm})^{2/81}
    \qty(\frac{\Sigma}{10\gm\cm^{-2}})^{-5/27}
        \nonumber\\
    &&\times 
    \qty(\frac{M}{10^{26}\gm})^{10/81}
    \qty(\frac{a}{1\au})^{-7/27}
        \nonumber\\
    &&\times 
    \fqty{\rho_{\rm p}}{2\gm\cm^{-3}}^{1/9}
    \fqty{\rho_{\rm gas}}{2\times 10^{-9}\gm\cm^{-3}}^{-2/27},
    \label{eq::separation_withgas}
\end{eqnarray}
and
\begin{eqnarray}
\langle \Tilde{e}^2\rangle ^{1/2} &\simeq& C_b^{-5/27} C_r^{25/27} m^{5/81} \Sigma^{1/27}
     M^{-2/81} a^{-4/27}
     \rho_{\rm p}^{1/9} \rho_{\rm gas}^{-5/27}\nonumber\\
     &\simeq&5.3\times 
     \qty(\frac{m}{10^{23}\gm})^{5/81}
    \qty(\frac{\Sigma}{10\gm\cm^{-2}})^{1/27}
        \nonumber\\
    &&\times 
    \qty(\frac{M}{10^{26}\gm})^{-2/81}
    \fqty{\rho_{\rm p}}{2\gm\cm^{-3}}^{1/9} 
        \nonumber\\
    &&\times 
    \qty(\frac{a}{1\au})^{-4/27}   
    \fqty{\rho_{\rm gas}}{2\times 10^{-9}\gm\cm^{-3}}^{-5/27},
    \label{eq::eccentricity_withgas}
\end{eqnarray} 
where
\begin{eqnarray}
    C_e &\equiv& \qty(\frac{6 \ln\Lambda}{C_{\rm D}\pi})^{1/5} \qty(3\pi^2)^{1/15},\\
    C_b &\equiv& \fqty{7}{\pi}^{1/5} \fqty{\pi}{3}^{1/15} .
\end{eqnarray}
According to these equations, $\Tilde{b}$ depends on $M$ and $\Sigma$.
Note that $\langle \Tilde{e}_m^2\rangle ^{1/2}$ also depends weakly on $m,\ \Sigma,\ M,$ and $a$.

Using eqs. \eqref{eq:isolationmass} and \eqref{eq::separation_withgas} , we obtain the isolation mass 
\begin{eqnarray}
    M_{\rm iso} 
    &\sim& C_M^{27/22}C_b^{75/44}C_e^{15/22} a^{141/44} M_\star^{-27/44} m^{1/22} 
            \nonumber\\
    &&\times 
\rho_{\rm p}^{9/44} \rho_{\rm gas}^{-3/22}
    \Sigma^{3/2}
    \nonumber\\
    &\simeq& 0.29M_\oplus
    \fqty{\Sigma}{10\gm\cm^{-2}}^{3/2}
    \fqty{m}{10^{23}\gm}^{1/22} 
    \nonumber\\ 
    &&\times 
    \fqty{M_\star}{M_\odot}^{-27/44}
    \fqty{\rho_{\rm p}}{2\gm\cm^{-3}}^{9/44} 
    \nonumber\\ 
    &&\times 
    \fqty{a}{1\au}^{141/44} 
    \fqty{\rho_{\rm gas}}{2\times 10^{-9}\gm\cm^{-3}}^{-3/22},
    \label{eq::isolationmass_withgas}
\end{eqnarray}
where
\begin{eqnarray}
    C_M &=& (2\pi)^{3/2}\qty(\frac{2}{3})^{1/2}.
\end{eqnarray}
The isolation mass depends mainly on the surface density $\Sigma$ and the semimajor axis $a$.

\section{Method of calculation}\label{sec:method}
We perform a series of {\em N}-body simulations of planetesimal accretion starting from a planetesimal ring around a central star.

\subsection{Plentesimal Ring Models}

We assume each planetesimal ring to be axisymmetric and with a radially uniform surface density around a solar-type ($1 M_\odot$) star.
In all the models, we take the center of the ring width to be located at $a_0=1.0 \au$, and we take the surface density $\Sigma$ of planetesimals to be
\begin{equation}
    \Sigma = 
    \begin{cases}
         \Sigma_0 & (a_0 -\frac{1}{2} w_{\rm init}<a< a_0 +\frac{1}{2} w_{\rm init})\\
        0 & (\rm otherwise)
    \end{cases},
\end{equation}
where $w_{\rm init}$ is the initial ring width.
We assume that all planetesimals have the same mass $m=1.257\times 10^{23}\gm \simeq 2.1\times 10^{-5}M_\oplus$.
We take the density of each planetesimal to be $\rho_{\rm p} = 2\gm\cm^{-3}$, so that the radius of each planetesimal is $\sim 250\km$.
We change the number of planetesimals $N_{\rm init}$ when we change the total mass so that the resolution of the simulations remains consistent.

The initial eccentricities and inclinations are given by Rayleigh distributions with RMS values 
$
\langle e^2\rangle^{1/2}
=2\langle i^2\rangle^{1/2}
=2 h_{m-m}
$,
where $h_{m-m}=(2m/3M_\odot)^{1/3}$ is the reduced mutual Hill radius of two planetesimals.
The argument of periapsis, the longitude of the ascending node, and the true anomaly are given randomly between 0 and $2\pi$.

Table \ref{table:ring_models} summarizes the initial parameters of the models.
In this paper, we refer to model W20M2 as the "fiducial model."
In the fiducial model, the total mass of planetesimals is $M_{\rm tot}=2.1 M_\oplus$, and the initial ring width is $0.2 \au$.
These values are the same as those of the planetesimal rings in \cite{Jason+2023Icar..39615497W}.
The total mass in the fiducial model is comparable to the sum of the terrestrial planets' mass in the solar system ($\sim 2 M_\oplus$), and it is also comparable to the mass of the planetesimal ring formed at around 1 au in the simulation by \citet{Izidoro+2022NatAs...6..357I} ($\sim 2.5 M_\oplus$).

\begin{deluxetable*}{Lccccccc}
\label{table:ring_models}
\tablecaption{Initial conditions of planetesimal rings}
\tablehead{
\colhead{Model name} 
& \colhead{$w_{\rm init}$ (au) } 
& \colhead{$M_{\rm tot}$ ($M_\oplus$) }
& \colhead{$N_{\rm init}$} 
& \colhead{$m$ (g)}
& \colhead{$\Sigma_0$ (g\,cm$^{-2}$)}
}
\startdata
    {\rm W02M2} & 0.025
    & $2.1$  
    & 100000 &$1.254\times 10^{23}$ & $340$\\
    {\rm W05M2} & 0.05   
    & $2.1$  
    & 100000 &$1.254\times 10^{23}$ & $170$\\
    {\rm W10M2}  & 0.1  & $2.1$ & 100000 & $1.254\times 10^{23}$ & $85$\\
    {\rm W20M2}  & 0.2  & $2.1$ & 100000 & $1.254\times 10^{23}$ & $42$\\
    {\rm W40M2}  & 0.4  & $2.1$ & 100000 & $1.254\times 10^{23}$ & $21$\\
    {\rm W80M2}  & 0.8  & $2.1$ & 100000 & $1.254\times 10^{23}$ & $11$\\
    {\rm W20M1}  & 0.2  & $1.05$ & 50000 & $1.254\times 10^{23}$ & $21$\\
    {\rm W20M4}  & 0.2  & $4.2$ & 200000  & $1.254\times 10^{23}$ & $85$\\
    \hline
\enddata
\end{deluxetable*}

We perform two runs for each model, using different random seeds to make the initial conditions.
In all models, we find that these two results are similar.
We therefore discribe the results from only one of the simulations, except when discussing the protoplanet distribution.

\subsection{Interaction with Disk Gas}\label{sec:gasdrag}

Particles in a gas disk feel the aerodynamic drag force exerted by the gas \citep[e.g.,][]{Adachi+1976PThPh..56.1756A}.
The drag force on a planetesimal, per unit mass, is given by \citep{Adachi+1976PThPh..56.1756A}
\begin{eqnarray}    
        \bm{F}_{\rm gas} &=& -\frac{1}{2m}C_{\rm D} \pi r_{\rm p}^2\rho_{\rm gas} |\bm{u}|\bm{u},
        \label{force_gasdrag}
\end{eqnarray}
where $m$ and $r_{\rm p}$ are the mass and radius of a planetesimal. 
In eq. \eqref{force_gasdrag}, $C_D$ is the drag coefficient, $\rho_{\rm gas}$ is the gas spatial density, and $\bm{u}=\bm{v}-\bm{v}_{\rm gas}$ is the velocity of the planetesimal relative to the gas.
{The gas velocity is described as $\bm{v}_{\rm gas} = (1-\eta) \bm{v}_\kep$ using the fractional difference $\eta$ between the Kepler velocity and the gas velocity and }
\begin{eqnarray}
        \eta &\equiv& -\frac{1}{2}\frac{c_{\rm s}^2}{v_\kep^2}\pdv{\ln P}{\ln r}
    \nonumber\\ 
    &=&
    -\frac{1}{2}\frac{c_{\rm s}^2}{v_\kep^2}\qty(\pdv{\ln \rho_{\rm gas}}{\ln r}+\pdv{\ln T}{\ln r}),\label{eq:eta}
\end{eqnarray}
{where $r$ is the distance from the star, $P$ is the gas pressure, and $c_{\rm s}$ is the sound speed.}
We adopt $C_D=2$ \citep{Adachi+1976PThPh..56.1756A}, and we assume the gas surface density to follow the simple power law
\begin{equation}
    \Sigma_{\rm gas} (r) = \Sigma_{\rm gas,0} \qty(\frac{r}{1\au})^{-1},
\end{equation}
where $\Sigma_{\rm gas,0}=1700\gm\cm^{-2}$ \citep{Jason+2023Icar..39615497W}.
The temperature profile is given by
\begin{equation}
    T = 2.8 \times 10^2 \qty( \frac{r}{1\au} )^{-1/2} \K,
\end{equation}
and the spatial density at the midplane is
\begin{equation}
    \rho_{\rm gas} = 0.99 \times 10^{-9} \qty( \frac{r}{1\au} )^{-9/4} \gm/\cm^3 .\label{method::gas_disk_profile}
\end{equation}
We use eq. \eqref{method::gas_disk_profile} for the gas density regardless of the height above the disk midplane because the planetesimals' inclination is small in our simulations.
{We neglect the gas disk dissipation for simplicity. The effect of gas dissipation is discussed in Sec. \ref{sec:summary}.}

Protoplanets orbiting in a gas disk change their orbits due to gravitational interactions with the gas disk, which results in eccentricity damping and orbital migration \citep[e.g.,][]{GoldreichTremaine_1979ApJ...233..857G,TanakaWard_TypeIEccentricity+2004ApJ...602..388T}.
For simplicity, {we neglect both the orbital migration and eccentricity and inclination damping due to the gravitational interactions with the gas disk.}
The effect of type-I migration on our results is discussed in Sec. \ref{sec:summary}.

\subsection{Numerical Method}

The planetesimals move under the influence of stellar gravity, mutual gravity, and gas drag.
The equation of motion for a planetesimal $j$ is
\begin{equation}
    \dv[2]{\bm{x}_j}{t}= - GM_\star\frac{\bm{x}_j}{|\bm{x}_j|^3} + \sum_{k=1, k\ne j}^N G m_k \frac{\bm{x}_k - \bm{x}_j}{|\bm{x}_k-\bm{x}_j|^3} + \bm{F}_{\rm gas} 
    ,
\end{equation}
where $m$ and $\bm{x}$ are the mass and position of a planetesimal.
For simplicity, we assume perfect accretion, in which all collisions between planetesimals lead to accretion, and we calculate planetesimal growth up to 2 Myr.

For the numerical integrations, we employ GPLUM \citep{GPLUM_2021PASJ...73..660I}, which uses the P$^3$T scheme \citep{P3T_2011PASJ...63..881O} with individual cutoff radius.
The code uses FDPS \citep[Framework for Developing Particle Simulator,][]{FDPS} for parallelization.
We perform simulations on the Cray XC50 supercomputer at the Center for Computational Astrophysics at the National Astronomical Observatory of Japan, and on Fugaku at RIKEN.
{We use up to 1152 CPU cores on Fugaku. The simulation takes approximately 500 hours for each run.}
\section{Results}\label{sec:results}
We first discuss in detail the results for the fiducial model W20M2 (which has $w_{\rm init}=0.2\au$ and $M_{\rm tot}=2.1M_\oplus$).
Then, we consider the dependence of the system characteristics on the initial ring conditions.

\subsection{Planetesimal Growth in the Fiducial Model}
\subsubsection{Overall Evolution}
\begin{figure*}    
  \centering
\begin{minipage}{0.45\textwidth}
\centering
  \includegraphics[width =\textwidth]{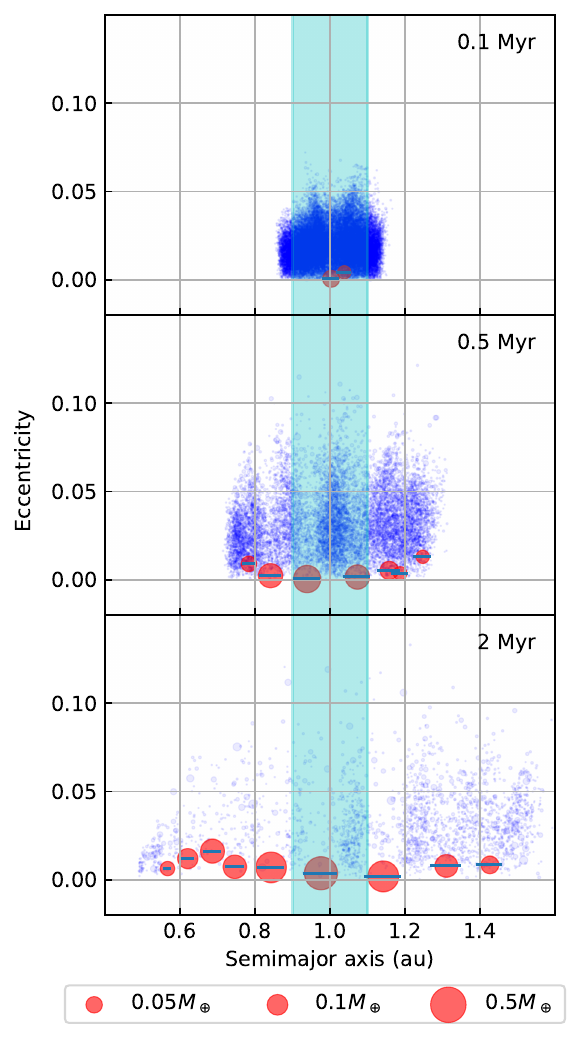}
\end{minipage}
\begin{minipage}{0.45\textwidth}
\centering
  \includegraphics[width=\textwidth]{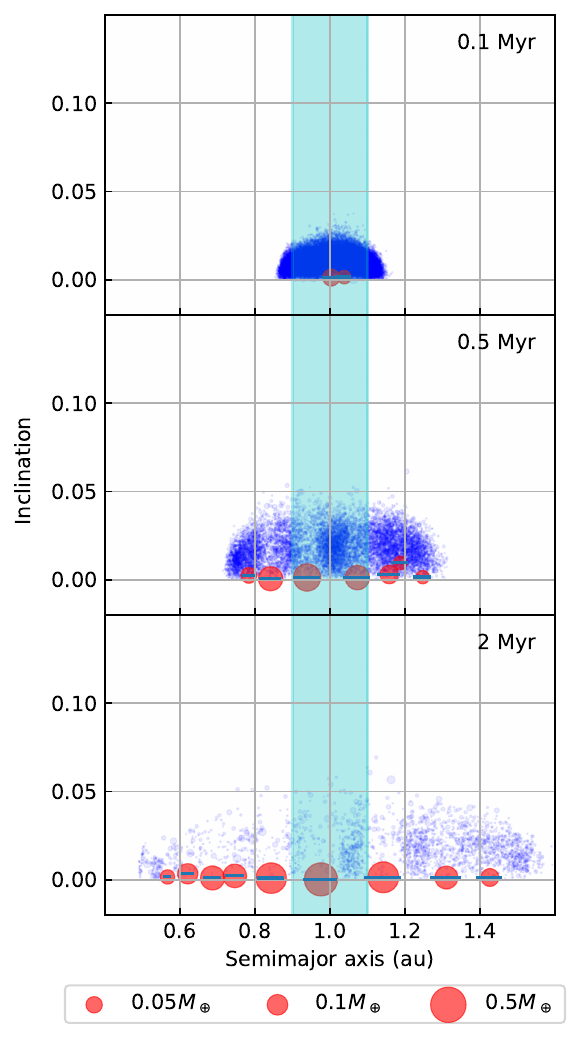}
\end{minipage}
\centering
  \caption{
  Snapshots of the fiducial-model simulation in (a) the $a$--$e$ and (b) the $a$--$i$ planes.
  The system initially consists of $10^5$ equal-mass ($m=1.257\times 10^{23}\gm$) planetesimals distributed in a narrow ring, with semimajor axes in the range 0.9--1.1 au.
  The blue dots represent planetesimals, and red circles represent protoplanets more massive than $1000m = 1.257\times 10^{26}\gm\simeq 0.02M_\oplus$.
  The ratios of the circles' radii correspond to the ratios of the particles' radii.
  The number of particles remaining is 33352 (at 0.1 Myr), 11466 (at 0.5 Myr), and 2343 (at 2 Myr).
  }
  \label{fig::fiducial::ae-ai}
\end{figure*}

Figure \ref{fig::fiducial::ae-ai} shows snapshots of the fiducial model in the $a$--$e$ and $a$--$i$ planes.
In this section, we define a protoplanet as a planetesimal with a mass greater than $1000m = 1.254\times 10^{26} \gm \simeq 0.021 M_\oplus$. 
Planetesimals begin runaway growth at first \citep{KI96+1996Icar..123..180K}.
Before the formation of protoplanets, the ring expands radially, due mainly to viscous stirring among the planetesimals, and the expansion rate is relatively small \citep{OhtsukiTanaka2003Icar..162...47O}.
At 0.1 Myr, two protoplanets have formed.
The protoplanets then undergo oligarchic growth while efficiently scattering the remaining planetesimals.
The ring thus expands faster than before protoplanet formation.
The growing protoplanets maintain their orbital separations $\gtrsim 10 r_{\rm H}$ due to orbital repulsion.
At the end of the simulation (2 Myr), the ring width is about 1.1\au, more than five times wider than its initial width (0.2\au).
The number of particles has decreased to 2343.
Nine protoplanets have foremd, with masses in the range 0.032--0.40 $M_\oplus$, and they contain $\sim$ 80\% of the total mass.

{Throughout this simulation, the eccentricities and inclinations of the protoplanets are smaller than those of most of the planetesimals because of dynamical friction.}
Dynamical friction remains effective in keeping the protoplanets' random velocities small, even at the end of the simulation.

\begin{figure}[ht]
  \centering
  \includegraphics[width=0.5 \textwidth]{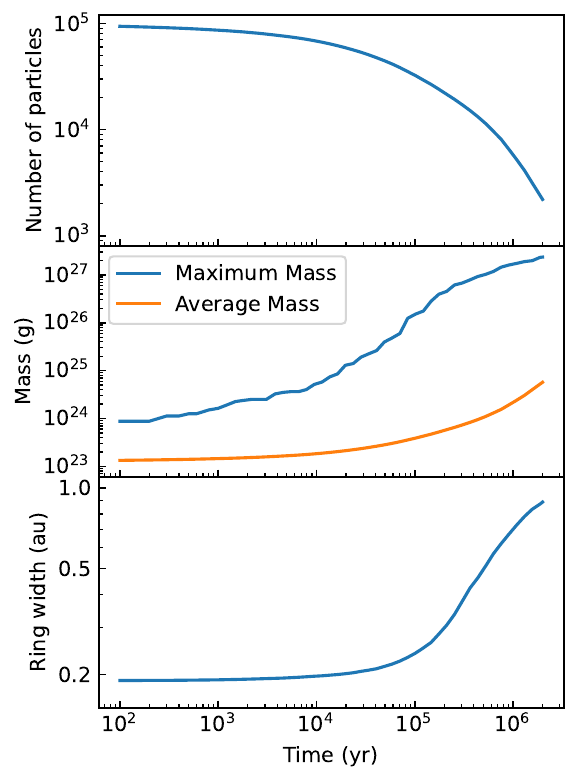}
  \caption{Time evolution of the number of particles (top), the mass of the 
  largest particle (middle, blue line), the average mass (middle, orange line), 
  and the ring width (bottom).
  }
  \label{fig::fiducial::nma}
\end{figure}

Figure \ref{fig::fiducial::nma} shows the evolution of the number, the maximum mass and the average mass of planetesimals, and of the ring width.
Here we define the "ring width" to be the width that contains 95\% of the total mass.
We define the inner semimajor axis $a_{\rm in}$ so that planetesimals with $a<a_{\rm in}$ have 2.5 \% of the total mass.
The outer semimajor axis $a_{\rm out}$ is defined in a similar way, so that 95\% of the total mass lies between $a_{\rm in}$ and $a_{\rm out}$.
As the system evolves, the number of particles decreases, and the mass of the most massive body increases.
The most massive body grows more rapidly than the average mass during the first 0.1 Myr of the system evolution; in other words, it initially experiences runaway growth.
After 0.1 Myr, however, the largest mass grows at about the same rate as the mean mass; that is, so the protoplanets grow in an orderly fashion, and the growth scheme transforms into oligarchic growth.
The ring width expands relatively slowly at first, but the expansion rate increases as the protoplanets grow because their scattering becomes more efficient.

\begin{figure}
  \centering
    \includegraphics[width=0.5 \textwidth]{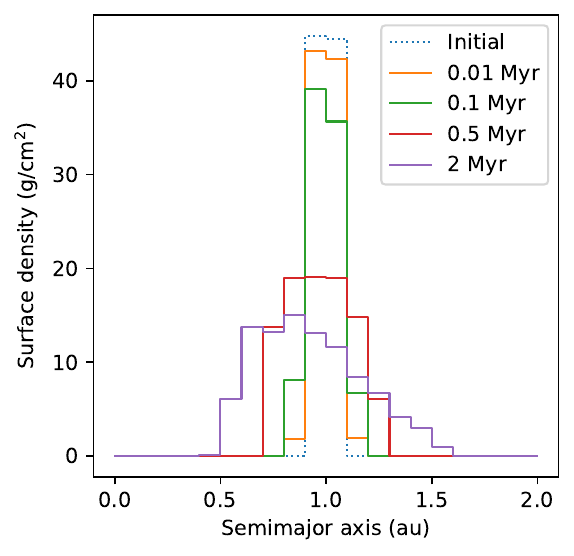}
  \caption{
  The surface density evolution in the fiducial simulation. 
  The blue dotted line shows the initial distribution.
  }
  \label{fig::fiducial::sd}
\end{figure}

We also investigate the surface density distribution of the particles.
We calculate this in the same way as for the minimum-mass solar nebula \citep{Weidenschilling_1977Ap&SS..51..153W, Hayashi_MMSN_1981PThPS..70...35H}.
We calculate the mass in each radial bin as follows:
(1) The mass of a planetesimal is added to the bin to which its semimajor axis belongs.
(2) The mass of protoplanet $j$ is allocated uniformly to the bins in the region ($a_{j,{\rm in}}, a_{j, {\rm out}})$, where $a_{j,{\rm in}} = \sqrt{a_{j-1}a_{j}}$ and $a_{j,{\rm out}} = \sqrt{a_{j}a_{j+1}}$.
For the innermost and outermost protoplanets, we set $a_{0,{\rm in}} = a_{0}- (a_{0,{\rm out}}-a_0)$ and $a_{n-1,{\rm out}} = a_{n-1}+ (a_{n-1}-a_{n-1,{\rm in}})$ , respectively.
Figure \ref{fig::fiducial::sd} shows the evolution of the surface density.
The wider the ring the lower its surface density becomes.
Moreover, the surface density retains a peak.

\begin{figure} [ht]
  \centering
  \includegraphics[width=0.5 \textwidth]{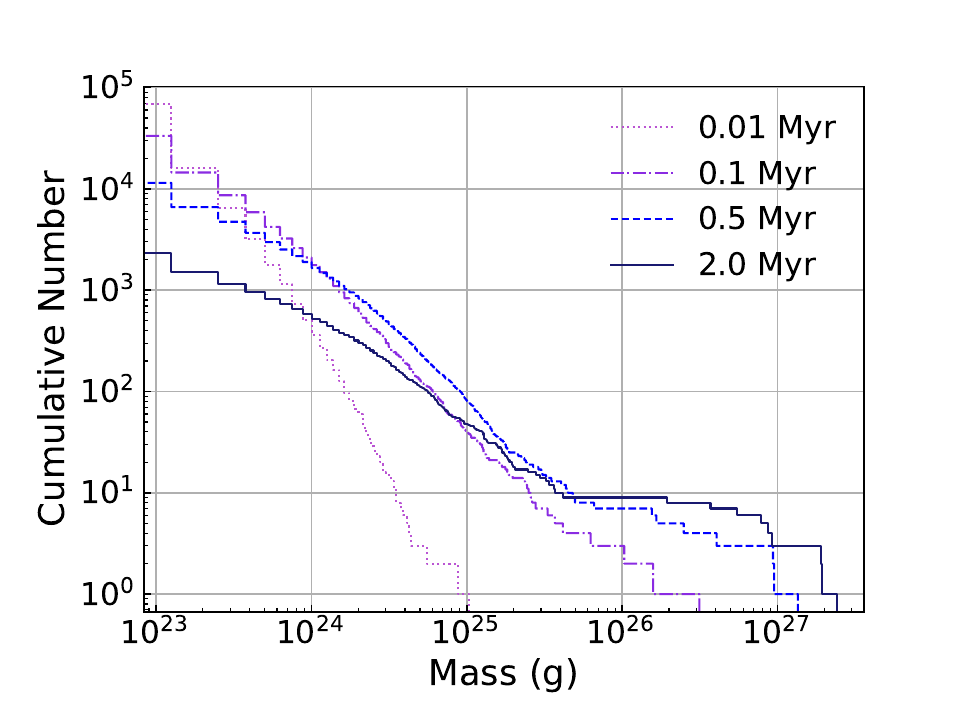}
  \caption{Cumulative number of particles plotted against mass at 0.01, 0.1, 0.5, 2.0\, Myr in the fiducial model.
  }
  \label{fig::fiducial::number}
\end{figure}

Figure \ref{fig::fiducial::number} shows the evolution of the planetesimal size distribution.
After 0.5\,Myr, nine protoplanets ($\gtrsim10^{26}\gm$) grow larger, leaving the other planetesimlas small.
This is typical for oligarchic growth \citep{KI02}.

\begin{figure} [ht]
  \centering
  \includegraphics[width=0.5 \textwidth]{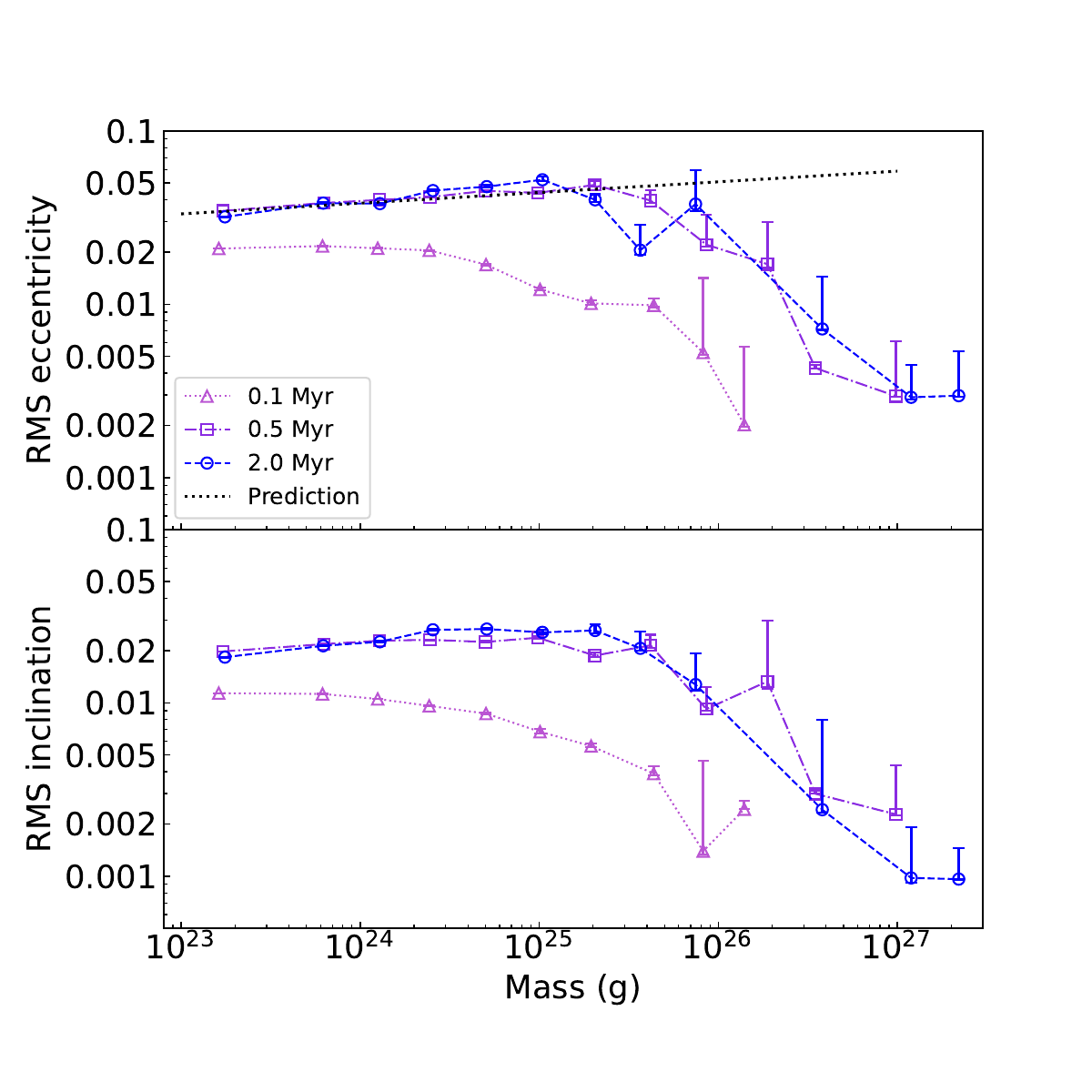}
    \caption{Evolution of the distribution of RMS eccentricity (top) and inclination (bottom) against mass at $t=0.01, 0.1, 0.5, 2.0\,{\rm Myr}$. 
    The dotted line in the top panel is the RMS eccentricity predicted by eq. \eqref{equilibllium_eccentricity} for $M=10^{27}\gm$ and $\Sigma = 10\gm\cm^{-2}$.
    The error bars represent the 68\% confidence interval of the RMS.}
  \label{fig::fiducial::randomvelocity}
\end{figure}

Figure \ref{fig::fiducial::randomvelocity} shows the evolution of the RMS eccentricity and inclination.
The dotted line in Fig. \ref{fig::fiducial::randomvelocity} shows a semi-analytic estimate from eq. \eqref{equilibllium_eccentricity} for $M=10^{27}\gm$ and $\Sigma = 10\gm\cm^{-2}$.
This shows that the eccentricity of the planetesimals is almost at the equilibrium state estimated by viscous stirring and gas drag, while that of the protoplanets ($\gtrsim 10^{26}\gm$) remains small ($\lesssim 0.01$) due to dynamical friction.
This feature also matches the standard oligarchic growth model well.

\subsubsection{Formation of protoplanets}

\begin{figure} [ht]
  \centering
  \includegraphics[width = 0.5\textwidth ]{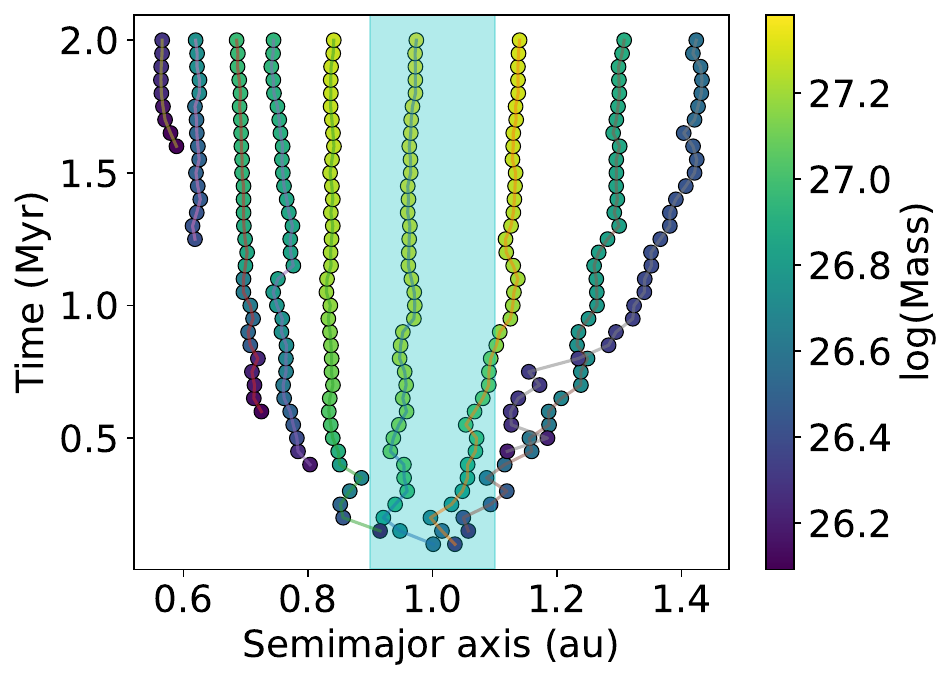}
  \caption{The evolution of the semimajor axes and masses of the protoplanets, which are more massive than $1000m$ at 2 Myr.
  Circles connected by a line represent the same particle.
  The cyan-shaded region represents the initial planetesimal distribution.
  }
  \label{fig::fiducial::migrationpath}
\end{figure}

Figure \ref{fig::fiducial::migrationpath} shows the growth paths of the protoplanets.
At first, planetesimals are distributed in a narrow region around 1 au. 
Before ring expansion, the planetesimal surface density is high, and the protoplanets grow efficiently at first.
As planetesimals are scattered and the ring expands, however, additional protoplanets start to grow away from the region of the initial ring.
As the ring expands, the surface density of the planetesimals decreases.
For this reason, protoplanets that began to grow earlier near the ring center can become more massive than those that grow away from the ring center.

\begin{figure} [ht]
  \centering
  \includegraphics[width = 0.5\textwidth ]{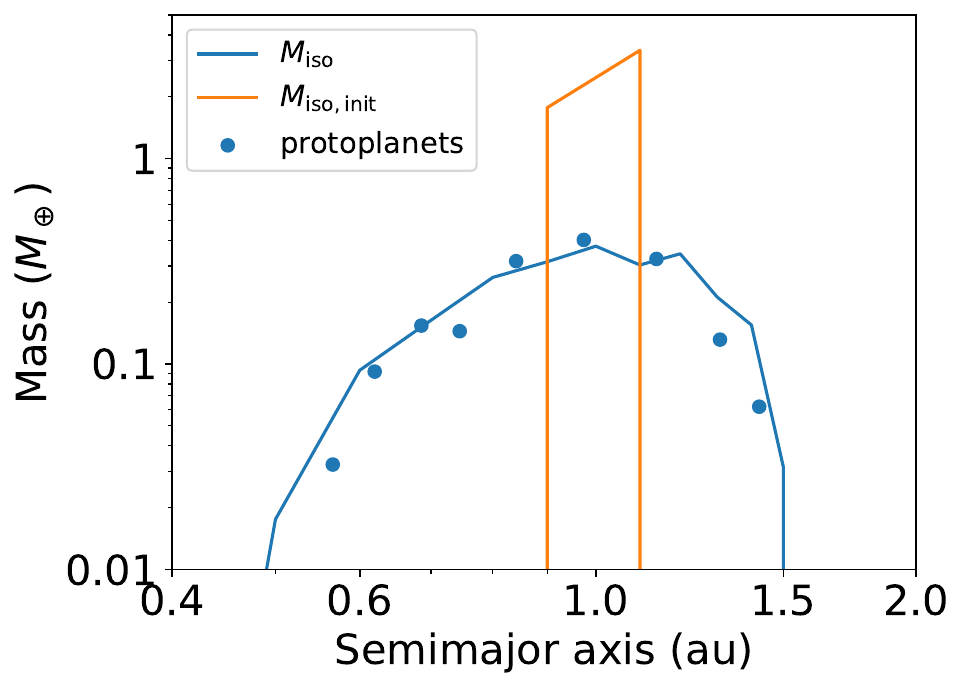}
  \caption{
  The distribution of protoplanets at 2 Myr in the $a$--$M$ plane.
  Each blue dot represents a protoplanet in the simulation.
  The solid line shows the local isolation mass calculated using eq. \eqref{eq::isolationmass_withgas} and the local surface density shown in Figure \ref{fig::fiducial::sd}.
  The orange dashed line shows the isolation mass calculated from the initial surface density $M_{\rm iso, init}$, neglecting planetesimal diffusion. 
  }
  \label{fig::fiducial::am_vs_miso}
\end{figure}

Figure \ref{fig::fiducial::am_vs_miso} shows the distribution of the protoplanets in the $a$--$M$ plane.
We also show the isolation mass calculated from eq. \eqref{eq::isolationmass_withgas} using the local surface density at the end of the simulation and using the initial surface density without planetesimal diffusion.
The masses of the protoplanets are one order of magnitude smaller than the isolation mass without planetesimal diffusion.
The distribution of the protoplanet masses in the simulation agrees well with the distribution of the isolation mass predicted from the diffused surface density.
We thus find that planetesimal diffusion significantly changes both the masses and distribution of protoplanets.

\begin{figure} [ht]
  \centering
  \includegraphics[width = 0.45 \textwidth]{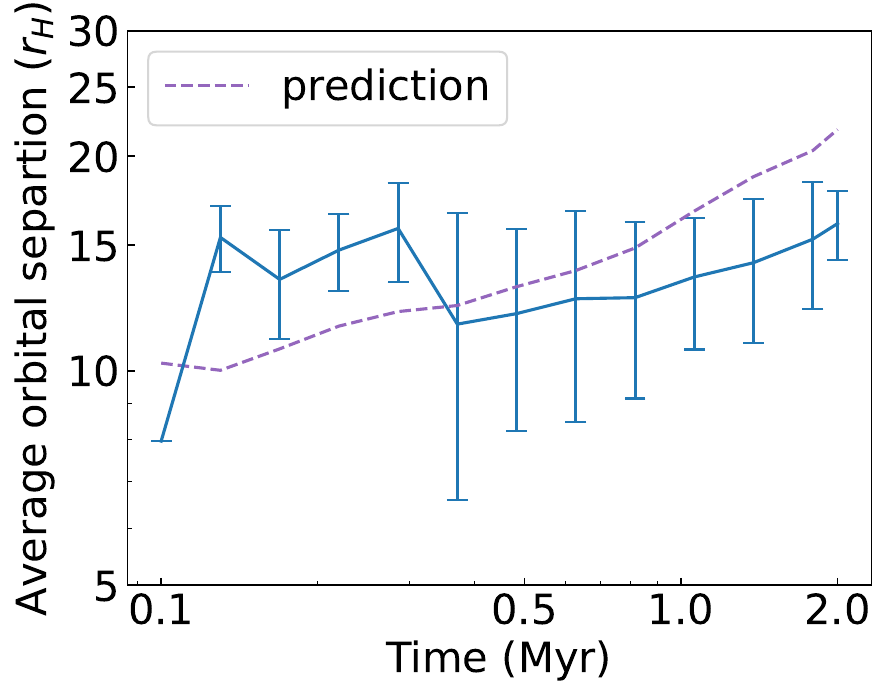}
  \caption{
  Evolution of the average orbital separation scaled by Hill radii. 
  The error bars show the standard deviations of the orbital separation at each time. 
  The purple dashed line shows the prediction from the oligarchic growth model using eq \eqref{eq::separation}.
   }
  \label{fig::fiducial::separation}
\end{figure}

Figure \ref{fig::fiducial::separation} shows the evolution of the average orbital separation of the protoplanets.
We also plot the orbital separation predicted by eq. \eqref{eq::separation} from the average protoplanet mass and planetesimal surface density in the simulation.
At first, the number of protoplanets is limited, so their orbital separations are relatively large.
After $\sim$ 0.2 Myr, the average orbital separation remains at $\simeq 10 r_{\rm H}$.
As the system evolves, the protoplanets become more massive, and the surface density of the planetesimals decreases; then, the orbital separation increases according to eq. \eqref{eq::separation_withgas}.
The gradual increase in the orbital separation is shown in Fig. \ref{fig::fiducial::separation}.
At the end of the simulation, the average orbital separation is slightly smaller than the predicted one.
This can be explained as follows:
Eq. \eqref{timescale::repelsion} was derived assuming a two-body encounter under solar gravity.
In reality, the repelling protoplanets also interact with neighboring protoplanets, which suppresses the expansion of the orbital separation.

The critical factor for oligarchic growth is the surface density of planetesimals.
Although the surface density distribution evolves in a planetesimal ring due to planetesimal diffusion, the simulation results agree well with the analytical estimates obtained from the oligarchic growth model.

\subsection{Parameter Dependence}
\subsubsection{Initial ring width}
\begin{figure}
  \centering
    \includegraphics[width=0.5 \textwidth]{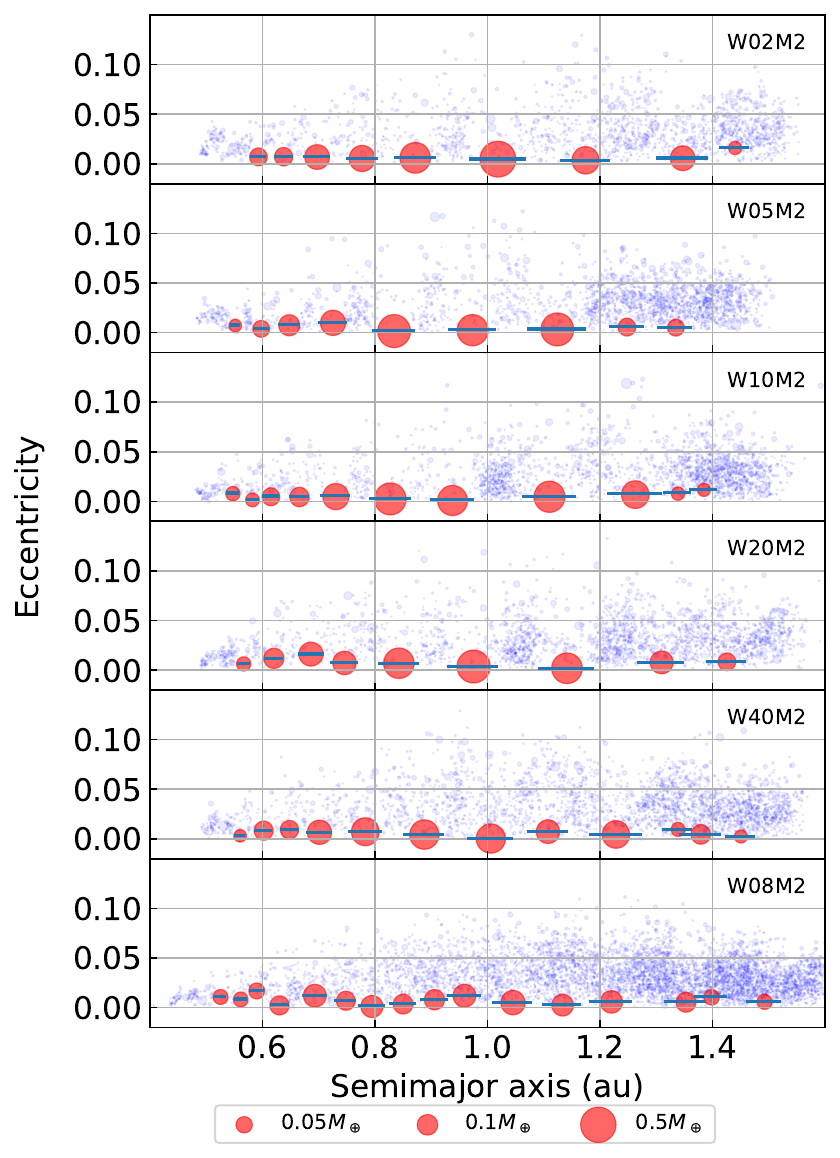}
  \caption{Snapshots of simulations in the $a$--$e$ plane.
  The blue dots represent planetesimals, and the red circles represent protoplanets more massive than $1000m = 1.257\times 10^{26}\gm\simeq 0.02M_\oplus$.
  The ratios of the radii of the circles correspond to the ratios of the radii of the particles.
  The numbers of particles remaining at this stage (2 Myr) are 1679 (W02M2), 2085 (W05M2), 2182 (W10M2), 2343 (W20M2), 2848 (W40M2), and 5851 (W80M2).
  }
  \label{fig::winit::ae}
\end{figure}

We next compare the results from six models: W02M2, W05M2, W10M2, W20M2, W40M2, and W80M2, where we vary the initial ring width while fixing the total mass.
Figure \ref{fig::winit::ae} shows snapshots of the simulations in the $a$--$e$ plane at the end of the simulation (2 Myr).
We find that the growth mode in each of the models is similar to that in the fiducial model; protoplanets undergo oligarchic growth while the ring width expands.
We compare the system properties at 2 Myr.
In all the simulations except for model W80M2, the protoplanets contain about 80\% of the ring mass.
In this sense, the systems are in the same evolutionary stage.
In model W80M2, the protoplanets contain only about 70\% of the ring mass.
This means that model W80M2 evolves more slowly than the other models.
This is natural because protoplanets grow faster in a denser environment, and model W80M2 has he lowest initial surface density of all the models.

\begin{figure}
  \centering
  \includegraphics[width=0.45 \textwidth]{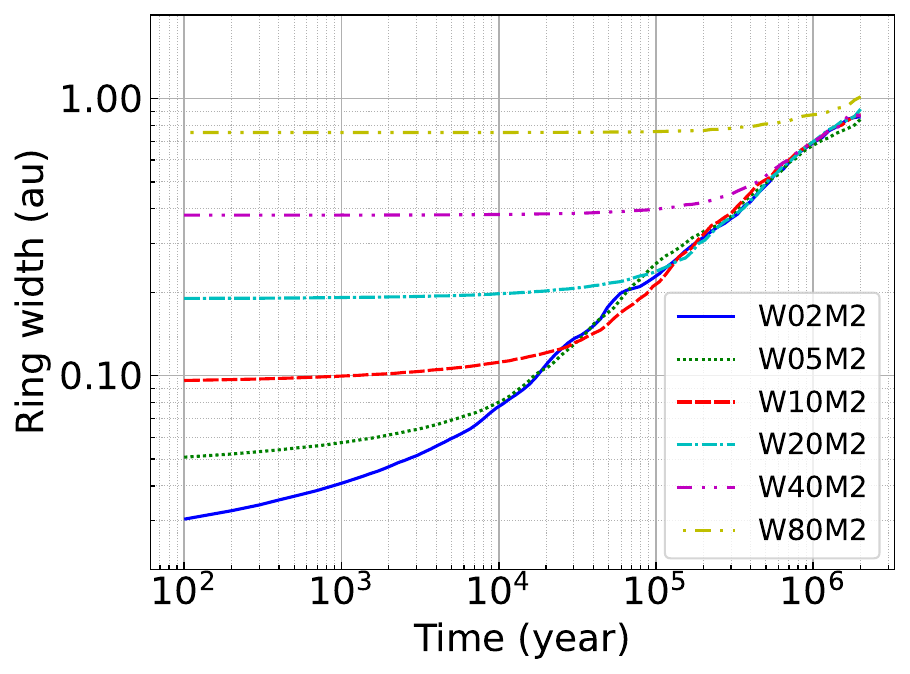}
  \caption{
  Evolution of the ring widths in simulations of models changing initial ring widths.
  The initial ring widths in the models are 0.025 au (W02M2), 0.05 au (W05M2), 0.1 au (W10M2), 0.2 au (W20M2), 0.4 au (W40M2) and 0.8 au (W80M2).
  }
  \label{fig::winit::ringwidth}
\end{figure}

In Figure \ref{fig::winit::ringwidth} we plot the evolution of the ring width, which we define to be the 95\% mass width.
When we fix the total mass, the narrower rings expand faster than the wider ones.
This occurs because protoplanets grow faster in the narrower rings, due to their higher surface density of planetesimals, which results in more efficient ring expansion than in the wider rings.
Regardless of its initial width, however, the ring width ultimately converges to a specific value.
In model W80M2, the ring width is slightly larger than that of the other rings.

\begin{figure} 
  \centering
  \includegraphics[width=0.45 \textwidth]{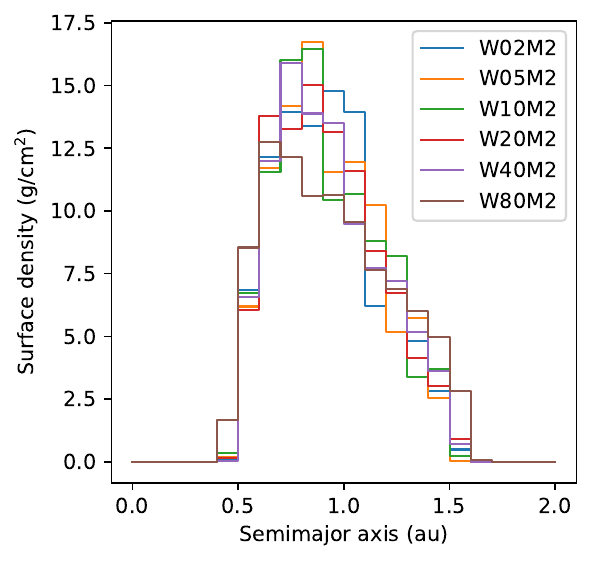}
  \caption{
  The surface density distribution at 2 Myr in models with different initial ring widths.
  }
  \label{fig::winit::sudfacedensity}
\end{figure}

Figure \ref{fig::winit::sudfacedensity} shows the surface density distribution at 2 Myr in models with different initial widths.
The surface density distributions exhibit little differences between the models, even though the initial ring widths differs by one order of magnitude.
This is consistent with the convergence of the ring widths to a specific value.

The characteristics of the protoplanets in each model are summarized in Table \ref{tab:winit:protoplanets}.
We can divide the models into two groups: the narrow rings ($w_{\rm init}\le 0.2 \au$) and the wide rings ($w_{\rm init}\ge 0.4 \au$).

In the case with the narrow rings, the number of protoplanets is typically $N_{\rm pp}\approx 10$.
The protoplanets' masses are in the range 0.02--0.53 $M_\oplus$, and the average orbital separation $\langle\Tilde{b}\rangle$ between adjacent protoplanets is 14--16 $r_\hill$.
These properties are almost independent of the initial width.
Conversely, in the wide rings, the number of protoplanets is slightly larger, and their masses decrease as the initial width increases.
The orbital separation also declines to $\sim 12 r_\hill$.

These results are explained by the differences in surface density.
In the narrow rings, the surface density distribution is almost independent of the initial width, whereas in the wide rings it decreases as the initial width increases.
Eq. \eqref{eq::isolationmass_withgas} shows that the masses of the protoplanets depend on the surface density.
The protoplanets gain similar masses in the narrow rings, while the masses are smaller in the wide rings.
If we fix the total mass, the number of protoplanets increases when they become less massive.
The orbital separations also depend on the surface density.
These dependencies are consistent with the protoplanet characteristics in the narrow and wide rings.

\begin{deluxetable*}{Lccccccc}
\label{tab:winit:protoplanets}
\tablecaption{Protoplanet properties in each simulation. The different lines for the same model show the results obtained by changing the random seeds to make the initial conditions.}
\tablehead{
\colhead{Model} & \colhead{$N_{\rm pp}$} & \colhead{$\langle\Tilde{b}\rangle$ ($r_{\rm H}$)} & \colhead{Mass range of protoplanets ($M_\oplus$)} & \colhead{Mass in protoplanets (\%)}}
\startdata
\text{W02M2} & 9  & 15.2 & 0.025-0.50 & 82 \\
\text{W02M2} & 9 & 15.5& 0.035-0.47 & 78 \\
\text{W05M2}  & 9 & 16.1& 0.022-0.39 & 76 \\
\text{W05M2}  & 11 & 13.9& 0.023-0.53 & 81 \\
\text{W10M2}   & 11 & 14.0& 0.024-0.36 & 80 \\
\text{W10M2}   & 9 & 15.6& 0.024-0.45 & 78 \\
\text{W20M2}  & 9 & 16.1& 0.032-0.40 & 79 \\
\text{W20M2}   & 10 & 15.7& 0.027-0.34 & 79 \\
\text{W40M2}   & 12 & 13.3& 0.022-0.28 & 77 \\
\text{W40M2}   & 11 & 13.2& 0.043-0.26 & 74 \\
\text{W80M2}   & 16 & 11.8& 0.024-0.18 & 67 \\
\text{W80M2}   & 16 & 12.6& 0.035-0.16 & 66 \\
\text{W20M1} & 8 & 14.8& 0.025-0.19 & 73 \\
\text{W20M1} &  9 & 13.0& 0.023-0.14 & 69 \\
\text{W20M4}  & 12 & 13.9& 0.028-0.78 & 81 \\
\text{W20M4}  & 11 & 15.6& 0.033-0.79 & 85 \\
\enddata
\end{deluxetable*}

\begin{figure} 
  \centering
  \includegraphics[width=0.45 \textwidth]{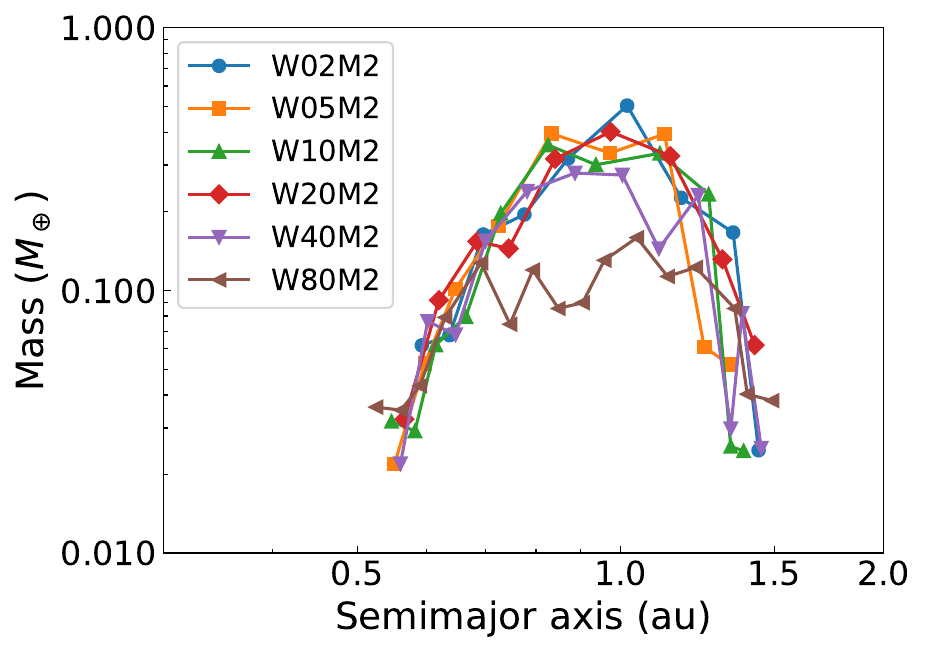}
  \caption{
  Masses versus the semimajor axes of the protoplanets at 2 Myr. 
  Each symbol represents a protoplanet in one of the simulations.
  Symbols with the same color and shape correspond to the same ring model.
  }
  \label{fig::winit::a-M}
\end{figure}

The distribution of the protoplanets in the $a$--$M$ plane is plotted in Figure \ref{fig::winit::a-M}.
As in the fiducial model, the protoplanets in all models are more massive near the region occupied by the initial ring than at the ring edge.
We also find that---similar to the surface density---the distribution of the protoplanets in the narrow rings does not depend strongly on the initial ring width.

\begin{figure} 
  \centering
  \includegraphics[width=0.45 \textwidth]{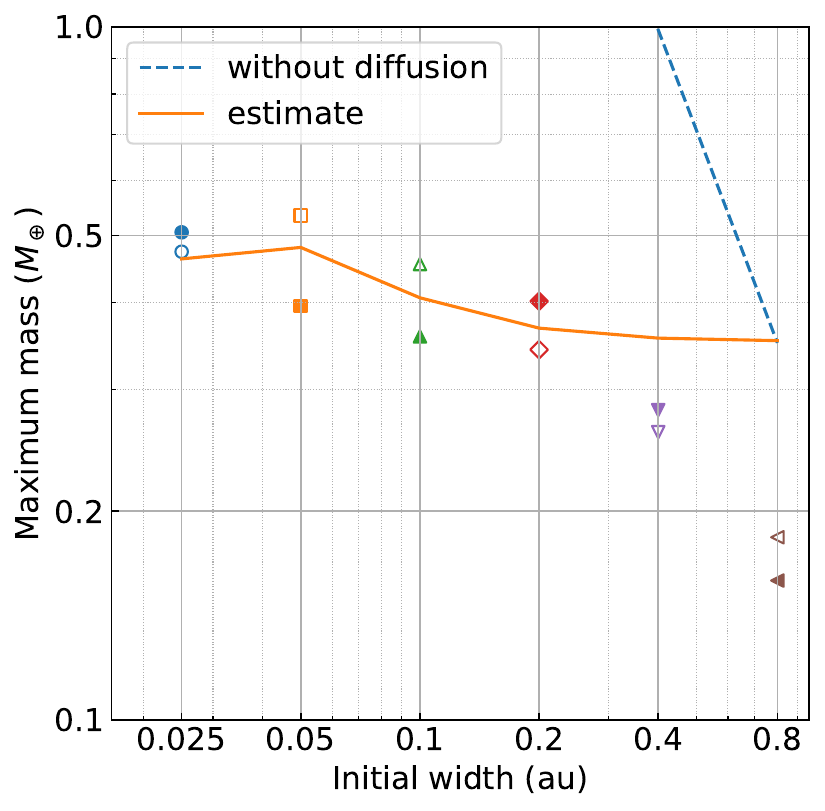}
  \caption{
  The largest protoplanet's mass in each simulation plotted against the initial ring width.
  The filled and open symbols correspond to different random seeds.
  The orange solid line corresponds to the maximum isolation mass from eq. \eqref{eq:isolationmass} using the local surface density in the simulation.
  The dashed blue line shows the isolation mass when planetesimal diffusion does not take place.
  }
  \label{fig::winit::winit_vs_maxmass}
\end{figure}

Figure \ref{fig::winit::winit_vs_maxmass} shows the maximum mass of the protoplanets in the models with different initial ring widths.
These masses are far smaller than the isolation mass obtained without planetesimal diffusion.
For an initial ring width smaller than 0.2 au, the largest protoplanet's mass is almost independent of the initial ring width.
This is also consistent with the surface density distribution.
Eq. \eqref{eq::isolationmass_withgas} shows that the isolation mass depends on the surface density, and we plot the maximum isolation mass obtained from the surface density in each simulation in Figure \ref{fig::winit::winit_vs_maxmass}.
In the narrow rings, the maximum protoplanet mass is comparable to the maximum isolation mass.
In the wide rings, however, the estimated isolation mass approaches the isolation mass without diffusion because planetesimal diffusion is not so effective in wider rings.
Moreover, the maximum protoplanet mass is smaller than the estimated isolation mass in the wide rings.
It is possible that protoplanet growth may be incomplete for the wide rings because the growth timescale is longer in environments with lower planetesimal surface densities. 
If this is the case, it is likely that protoplanet growth and planetesimal diffusion will continue after 2 Myr and bring the protoplanet mass into agreement with the estimated isolation mass.

\subsubsection{Dependence on the total mass}
\begin{figure}[ht]
  \centering
    \includegraphics[width=0.5 \textwidth]{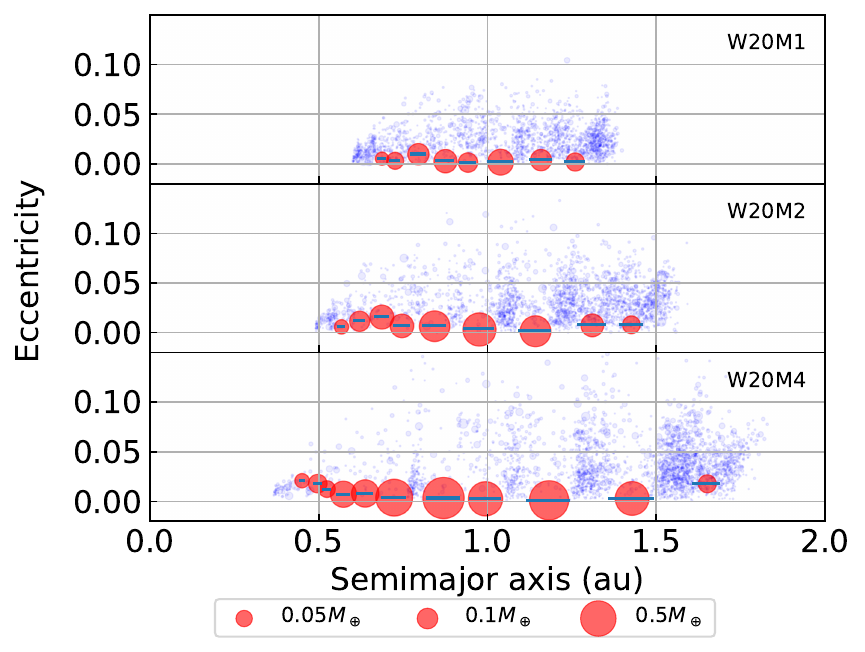}
  \caption{ 
  The same as Fig. \ref{fig::winit::ae}, but for models with different total ring masses.
  The numbers of particles remaining at this stage (2 Myr) are 1929 (W20M1), 2343 (W20M2), and 2705 (W20M4).
  }
  \label{fig::mtot::ae}
\end{figure}

We next compare the results from three models---W20M1, W20M2, and W20M4---where we vary the total mass while fixing the initial ring width.
We find that the growth mode in each model is similar to that in the fiducial model: protoplanets undergo oligarchic growth while the ring width expands.
Figure \ref{fig::mtot::ae} shows the snapshots of these systems in the $a$--$e$ plane at the end of the simulation (2 Myr).
In all calculations, the protoplanets contain about 75\%--85\% of the total mass at 2\,Myr.
In this sense, all the systems are at almost the same evolutionary stage at 2\,Myr.

\begin{figure} 
  \centering
  \includegraphics[width=0.45 \textwidth]{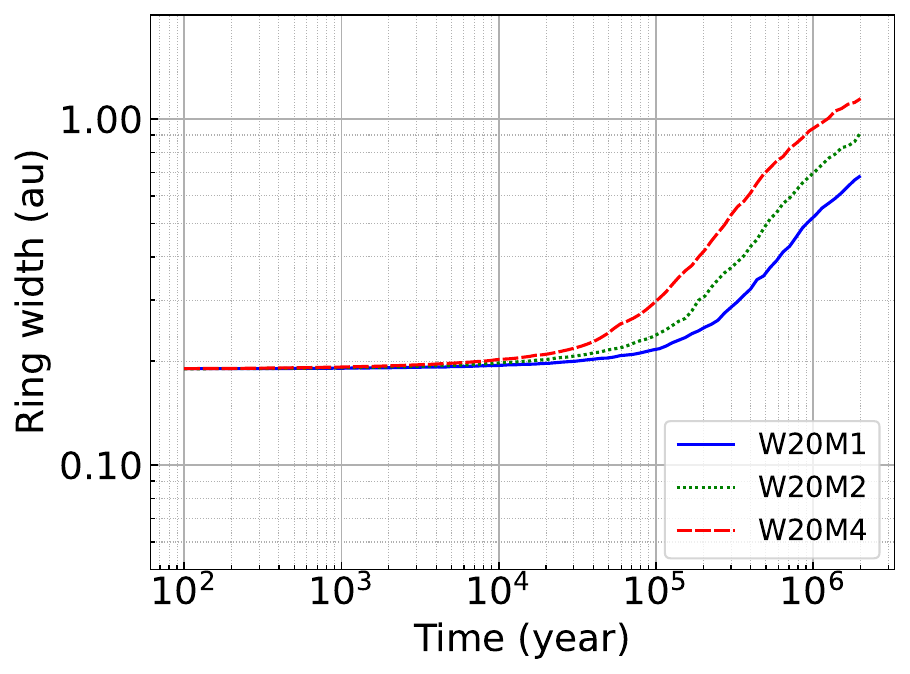}
  \caption{
  The same as Fig. \ref{fig::winit::ringwidth}, but with different total ring masses.
  The total masses are $1.05M_\oplus$ (W20M1), $2.1 M_\oplus$ (W20M2), and $4.2 M_\oplus$ (W40M2).
  In each case, the initial ring width is 0.2 au.
  }
  \label{fig::mtot::ringwidth}
\end{figure}

Figure \ref{fig::mtot::ringwidth} shows the time evolution of the ring width.
In a more massive ring, protoplanets grow faster than in a less massive ring due to the higher surface density.
The larger protoplanets drive faster ring expansion, which results in faster expansion of a more massive ring than of a less massive one, as seen in the simulation.

\begin{figure} 
  \centering
  \includegraphics[width=0.45 \textwidth]{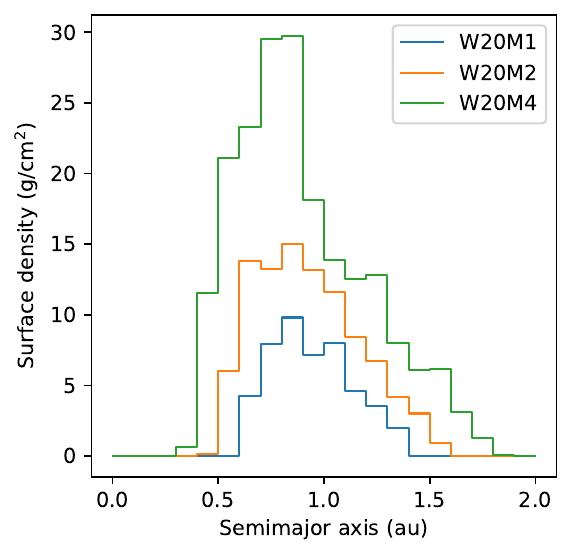}
  \caption{
  The same as Fig. \ref{fig::winit::sudfacedensity}, but with different total mass.
  }
  \label{fig::mtot::sudfacedensity}
\end{figure}

Figure \ref{fig::mtot::sudfacedensity} shows the surface density distribution in each ring.
Although the more massive ring expands faster than the less massive rings, its surface density remains larger than those of the less massive rings.

The characteristics of the protoplanets in each model are summarized in Table \ref{tab:winit:protoplanets}.
In most cases, there are about 10 protoplanets.
The maximum mass of the protoplanets are larger in the more massive rings, and their average orbital separations are 13--15 $r_\hill$.

\begin{figure} 
  \centering
  \includegraphics[width=0.45 \textwidth]{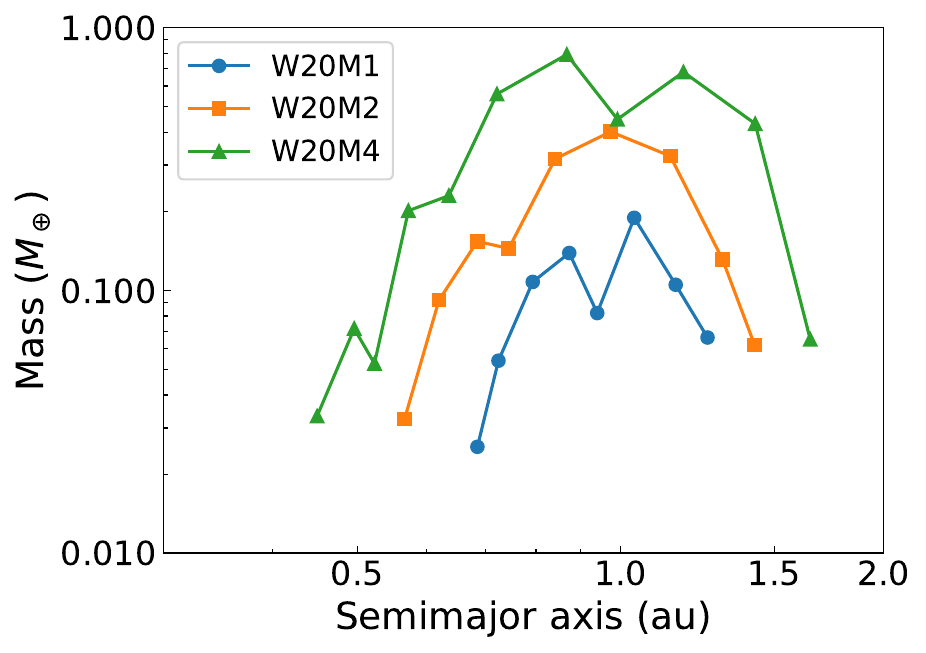}
  \caption{
  The same as Fig. \ref{fig::winit::a-M}, but for models with different total masses.
  }
  \label{fig::mtot::a-M}
\end{figure}

Figure \ref{fig::mtot::a-M} shows the protoplanet distribution in the $a$--$M$ plane.
In all the models, the protoplanets are less massive near the edge of the ring.
Moreover, at the same semimajor axis, the protoplanet mass is larger in a more massive ring than in a less massive ring.
This results from the higher surface density in the more massive ring.

\begin{figure} 
  \centering
  \includegraphics[width=0.45 \textwidth]{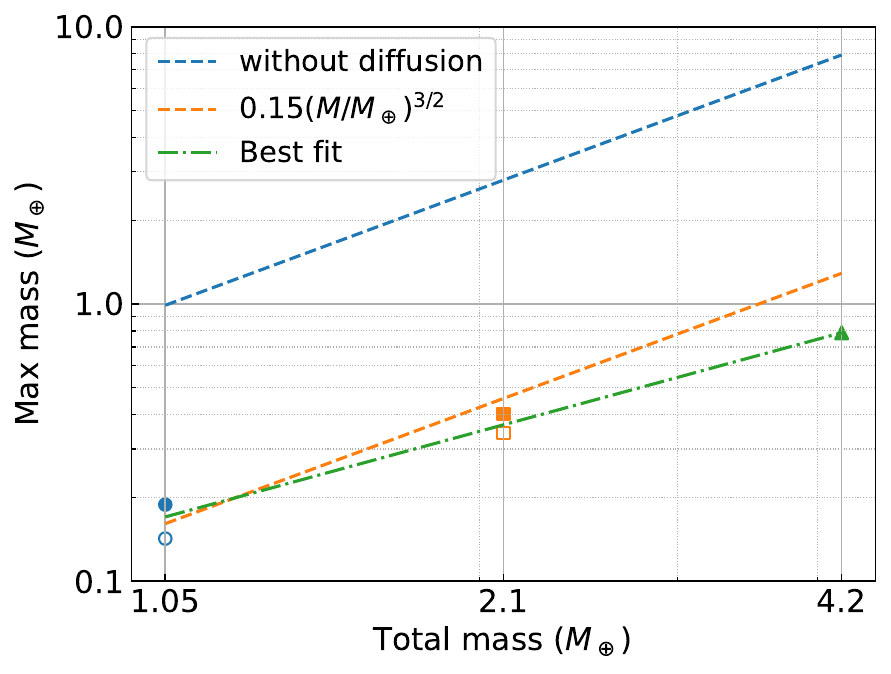}
  \caption{
  The mass of the largest protoplanet in each simulation plotted against the total mass.
  The filled and open symbols correspond to different random seeds.
  }
  \label{fig::mtot::mtot_vs_maxmass}
\end{figure}

The maximum mass of the protoplanets are plotted in Figure \ref{fig::mtot::mtot_vs_maxmass}, together with an empirical fit to the data.
Without radial diffusion, the isolation mass is proportional to $\Sigma^{3/2} \propto M_{\rm tot}^{3/2} w_{\rm init}^{-3/2}$.
In the best-fit model, the maximum mass of a protoplanet is $\propto M_{\rm tot}^{1.10\pm 0.06}$; this is a slightly weaker dependence on the total mass than in the model without planetesimal diffusion.
This is explained by the faster diffusion in a more massive ring, which results in a smaller surface density of planetesimals.

\section{Summary and discussion}\label{sec:summary}

In this work, we have investigated the planetesimal accretion in an expanding planetesimal ring using {\em N}-body simulation.
We changed the initial ring width and total mass systematically and investigated the dependence of the system structure on the initial conditions.
Our main findings are the following:
\begin{enumerate}
    \item 
        In a planetesimal ring, protoplanets undergo oligarchic growth while the ring expands.
        Viscous stirring by the planetesimals causes the initial ring expansion.
        After protoplanet formation, ring expansion is driven by the orbital repulsion between the protoplanets and by the scattering of planetesimals by the protoplanets.
        Rings expand more efficiently after protoplanet formation than before.
    \item
        We have confirmed that the oligarchic growth model holds even in a diffusing planetesimal ring.
        The isolation mass and orbital separation are well explained by eqs. \eqref{eq:isolationmass} and \eqref{eq::separation_withgas}, respectively, using the local planetesimal surface density. 
        Massive protoplanets are formed near the center of the ring because they grow in a denser region before planetesimal diffusion.
        Protoplanets also form outside the region occupied by the initial ring because the ring expands radially.
        Their masses are relatively small due to the small surface density of the planetesimals in the outlying locations.
    \item 
        When the total mass of planetesimals is fixed, the ring width expands faster when its initial width is narrower, and the ring widths ultimately take similar values, independent of the initial ring width. 
        If the initial width is sufficiently narrow, the mass and the orbital distribution of the protoplanets depend  weakly on the initial width.
    \item 
        When the initial ring width is fixed, the larger the total mass, the more massive do the protoplanets become, and the faster the ring expands.
        This is due to the faster planetesimal diffusion in a more massive ring. 
\end{enumerate}

In a planetesimal ring, the planetesimal surface density changes as the planetesimals diffuse.
Because the protoplanet properties are determined by the planetesimal surface density in the oligarchic growth model, it is vital to understand planetesimal diffusion in order to predict them.
{Radial diffusion of a planetesimal ring is caused by angular momentum transport originated from interaction between the particles such as gravity and collisions \citep[e.g., ][]{GoldreichTremaine1978a_1978Icar...34..227G,Tanaka+2003Icar..161..144T,OhtsukiTanaka2003Icar..162...47O}. 
\cite{Tanaka+2003Icar..161..144T} formulated the angular momentum transport for a swarm of planetesimals, and \cite{OhtsukiTanaka2003Icar..162...47O} evaluated the planetesimal diffusion timescale in an equal-mass system.
However, evaluating the angular momentum transport of a system with a mass-spectrum and growing protoplanets is far more complicated.}

In the present simulations, we have made some simplifications.
First, we assumed the surface density of planetesimals to be uniform in the radial direction.
{Although the initial planetesimal surface density is dependent on many parameters such as gas surface density and planetesimal formation efficiency \citep[][]{Izidoro+2022NatAs...6..357I,Morbidelli+22,Hyodo+2022A&A...660A.117H},}
we expect the basic processes discussed above are likely to be independent of the initial distribution of planetesimals; that is, planetesimal rings are likely to expand and protoplanets to undergo oligarchic growth in the expanding ring.
However, the masses and the orbital distribution of the protoplanets may be different when we change the initial surface density distribution in the ring; the dependence on the initial distribution remains to be studied in a future work.

The second assumption is that the gas surface density follows a simple power law, as in eq. \eqref{method::gas_disk_profile}.
When planetesimals are formed in a narrow ring, a gas pressure bump is likely to exist around the inner edge of the planetesimal ring \citep{Izidoro+2022NatAs...6..357I,Morbidelli+22}.
{Changing the gas disk model alters the gas drag force. 
Since the gas drag causes random velocity damping and orbital migration of planetesimals, it may be important to consider a gas pressure bump rather than a simple power-law profile. 
However, our simplification barely affects the simulation results. 
As shown in eq. \eqref{force_gasdrag}, the gas drag force is proportional to the square of the relative velocity u between a planetesimal and the gas. The relative velocity $\bm{u}$ depends on $\eta$ defined in eq. \eqref{eq:eta} and planetesimals’ random velocity $e, i$. A different gas profile leads to a different value of $\eta$.
However, when $e, i \gg \eta$, the relative velocity is mainly determined by $e$ and $i$ rather than $\eta$ \cite[e.g.,][]{Adachi+1976PThPh..56.1756A}. 
In our setup, $\eta \simeq 0.003$ and typically $e, i \gtrsim 0.01$ for planetesimals, so the condition $e, i \gg \eta$ is fulfilled during the simulation. 
Therefore, the gas drag force depends on $e$ and $i$ rather than $\eta$. 
We expect that our results barely change even when we adopt a different gas disk structure.
}
{Moreover, the migration timescale due to gas drag is $\sim 1\text{--}10$ Myr \citep{Adachi+1976PThPh..56.1756A}, which is far longer than the growth timescale for $\sim 100$-km-sized planetesimals.
Therefore, a different structure for the gas disk, which results in a different value of $\eta$, is not likely to alter our results in any essential way.}

{
We also assumed that the gas disk does not dissipate.
As it dissipates, the drag force decreases as in eq. \eqref{force_gasdrag}.
However, as shown in eqs. \eqref{eq::separation_withgas}, \eqref{eq::eccentricity_withgas}, and \eqref{eq::isolationmass_withgas}, the equilibrium eccentricity, the orbital separation, and the isolation mass do not depend strongly on $\rho_{\rm gas}$.
Therefore, the steady gas disk approximation may not change our results in any essential way.
}

As noted previously, we have neglected type-I migration for {simplicity} in the present study.
When this is taken into account, the gas disk profile decisively affects the migration of the protoplanets \citep[e.g.,][]{Ogihara_SE_2018A&A...615A..63O,Ogihara_terrestrial+2018A&A...612L...5O}.
However, we anticipate that the growth mode will still be oligarchic, as seen in the radially expanding rings.
To investigate the effects of type-I migration on the masses and orbital distributions of the protoplanets will require further simulations that include with type-I migration and the gas disk structure.

{In the present study, we ignore the accretion of pebbles.
The accretion of pebbles may cause planetary embryos to grow more efficiently than by planetesimal accretion \citep[e.g.,][]{Ormel&Klahr_2010A&A...520A..43O, Johansen+2021SciA....7..444J}, although the growth rate of a planet due to pebble accretion depends strongly on the pebble flux onto the star \citep{Lambrechts+2019A&A...627A..83L}.
Conversely, this contribution may not be as important as the accretion of planetesimals for rocky planet formation \citep{Izidoro+21_Planetesimal_vs_Pebble_2021ApJ...915...62I,Izidoro+2022NatAs...6..357I,Batygin&Morbidelli_2023NatAs...7..330B,Morbidelli+25_terrestrialplanets_pebble_or_planetesimal_2025E&PSL.65019120M,Shibata_Izidoro_2025ApJ...979L..23S}.
Which of these two is the dominant process in planetary growth is still under debate; this contribution can be studied in future work.}

The present work is also limited to terrestrial planet formation around a solar-type star.
However, a planetesimal ring can also be formed beyond the snow line \citep{Morbidelli+22,Izidoro+2022NatAs...6..357I}. 
\cite{KobayashiTanaka_2021ApJ...922...16K} performed a one-dimentional simulation of the solid particle growth from dust to planets, and they showed that the continuous formation of planetesimals around 5 au enables the rapid formation of giant planet cores.
Further simulations in the giant planet region are therefore needed to understand fully the planet formation process from planetesimal rings.

We thank referee André Izidoro  for valuable and constructive comments. 
This work was supported by MEXT as “Program for Promoting Researches on the Supercomputer Fugaku” (Structure and Evolution of the Universe Unraveled by Fusion of Simulation and AI; Grant Number JPMXP1020230406) and used computational resources of the supercomputer Fugaku provided by the RIKEN Center for Computational Science (Project ID: hp230204, hp240219).
Numerical computations were in part carried out on Cray XC50 at the Center for Computational Astrophysics, National Astronomical Observatory of Japan. 
Y.K. is supported by the Forefront Physics and Mathematics Program to Drive Transformation (FoPM), a World-leading Innovative Graduate Study (WINGS) Program at the University of Tokyo and by JST SPRING, Grant Number JPMJSP2108.
E.K. is supported by JSPS KAKENHI Grants No.\ 18H05438, 24K00698, and 24H00017.

%




\bibliography{bib}{}
\bibliographystyle{aasjournal}



\end{document}